\RequirePackage{fix-cm}
\documentclass[smallextended]{svjour3}       
\smartqed  
\usepackage{mathptmx}      
\usepackage{balance}
\usepackage{cite}

\usepackage{graphicx}

\usepackage{amsmath}
\usepackage{optidef}

\usepackage[table,xcdraw]{xcolor}
\usepackage{multirow}
\usepackage{paralist}

\usepackage{url}
\usepackage{pdfpages}

\usepackage{hyphenat}
\usepackage{xcolor}
\definecolor{added}{rgb}{0.65, 0.25, 0.39}
\newcommand{\added}[1]{#1} 

\journalname{Multimedia Tools and Aplications}

\begin{document}

\title{A Perceptual Measure for Evaluating the Resynthesis of Automatic Music Transcriptions}

\titlerunning{A Perceptual Measure for Resynthesis of AMT}        

\author{Federico Simonetta         \and
	Federico Avanzini \and
	Stavros Ntalampiras 
}


\institute{All authors are with LIM -- Music Informatics Laboratory,\\
	Department of Computer Science,\\
	University of Milano\\
	Email: \{name\}.\{surname\}@unimi.it
}

\date{Received: date / Accepted: date}

\maketitle

\begin{abstract}
	This study focuses on the perception of music performances when contextual factors, such as room acoustics and instrument, change. We propose to distinguish the concept of ``performance'' from the one of ``interpretation'', which expresses the ``artistic intention''. Towards assessing this distinction, we carried out an experimental evaluation where \added{91} subjects were invited to listen to various audio recordings created by resynthesizing MIDI data obtained through Automatic Music Transcription (AMT) systems and a sensorized acoustic piano. During the resynthesis, we simulated different contexts and asked listeners to evaluate how much the interpretation changes when the context changes.
	Results show that: (1) MIDI format alone is not able to completely grasp the artistic intention of a music performance; (2) usual objective evaluation measures based on MIDI data present low correlations with the average subjective evaluation. To bridge this gap, we propose a novel measure which is meaningfully correlated with the outcome of the tests.
	In addition, we investigate multimodal machine learning by providing a new score-informed AMT method and propose an approximation algorithm for the $p$-dispersion problem.
\end{abstract}

\section{Introduction}
\label{sec:introduction}

Automatic Music Transcription (AMT) can be broadly defined as the process elaborating on digital audio recordings in order to infer a specific set of relevant musical parameters of the sounds, and to convert them in some form of notation\cite{klapuri2004automatic}. Nowadays, AMT is a broad signal processing field encompassing a wide gamut of tasks and approaches. As an example, the output of an AMT system can be a traditional score, a Standard MIDI File (SMF), or a set of ad-hoc features\cite{benetos201901automatic}. A traditional score is a sequence of symbols that describes music according to the western notation and focuses on expressing music in a human-readable way so that it can be easily reproduced. SMFs instead describe the performance itself, possibly sacrificing precision at the semantic level (which is useful to the musician for performing the piece) while gaining precision in the description of the physical events that happened during the execution -- i.e. velocity and duration with which the keyboard keys were pressed, the pedaling timing, etc.

SMFs originate from the description of keyboard music and it hardly adapts to other instruments; moreover, the constantly increasing importance of Music Information Retrieval (MIR) created the need for a different symbolic representation of music sounds: hence, several AMT systems extract MIR features -- e.g. $f0$ estimation, intensity levels, and timbral descriptors\cite{rizzi2019instrument,fu2011Trans.}. The input to AMT systems is a variable itself: most authors focus on mono-modal methods which take as input only the audio recordings, while other methods tackle the problem with multimodal approaches\cite{simonetta201901multimodal} such as audio and scores (score-informed)\cite{jeong2020noteintensity,devaney2017evaluation,akbari2015piano,wang201710identifying}.

Regarding the resynthesis of MIDI transcribed recordings, many studies have shown that performers change their way of playing according to contextual conditions, such as physical properties of the instrument, reverberation and room acoustics, often even unconsciously\cite{kalkandjiev2015influence}. Recently, the authors of \cite{xu201905transferring} proposed a method to automatically transfer a piano performance across different contexts (instruments and environments) in order to make the reproduced sound as similar as possible to the original one, by adapting MIDI velocities and duration to the new context. However, they assume knowledge of the original piano parameters to carry out the adaptation; moreover, they use the same microphones and post-processing pipelines in every different context.

There are very few attempts addressing the subjective evaluation of AMT systems, with the notable exception of \cite{ycart2020investigating}. The authors prepared more than 150 questions asking subjects to chose the best transcription of a reference audio clip lasting 5-10 seconds and managed to collect 4 answers per question. We use the results of the above work as a main reference for the present study. 

We propose a methodology for evaluating the resynthesis of MIDI recordings extracted through AMT systems, taking into account contextual conditions in the resynthesis. Specifically, we wish to adapt the performance to the new resynthesized context while having knowledge only of the target context and of the original recording. In doing so, we propose a conceptual framework which distinguishes between the actual performance and the underlying artistic intention (i.e., the interpretation), and we design a methodology assessing to which extent such interpretation is perceivable in the resynthesized recording. \added{Possible applications that may be impacted by the proposed study are manifolds. The long-term objective of this study is the resynthesis of music for production
	and restoration purposes -- see Section~\ref{sec:concepts}. Other possible
	use-case scenarios include musicological studies and music teaching applications, such as
	the analysis and comparison of the interpretation, which in turns can pave completely new
	paths for these research fields; moreover, the ability of transcribing both
	performance and interpretation would also allow the comparison of the manifold
	ways in which different performers adapt their interpretation. Not least,
	architectural studies could be impacted from robust context-aware AMT models.}

The contributions of our work are \begin{inparaenum}
	\item an indication of MIDI format's inability to completely capture the artistic intention of a music performance and \item a perceptually-motivated measure for the evaluation of AMTs\end{inparaenum}. In addition, we investigate multimodal machine learning technologies applied to AMT by providing a new score-informed method and propose an approximation algorithm for the $p$-dispersion problem to optimally-chose the excerpts for the test.

For the purpose of  comparability and reproducibility of the results, the code is made available online\footnote{\url{https://github.com/LIMUNIMI/PerceptualEvaluation}} \added{and the full set of the computed statistics are available in the Supplementary Materials}.

\section{Restoration, Performance and Interpretation}
\label{sec:concepts}

One of the long-term motivations behind the present work is \added{the automatic restoration of old and contemporary} music recordings by reproducing the performances as accurately as possible. \added{If audio restoration
	processes~\cite{godsill1998digital} usually target old and deteriorated operas, restoration of
	contemporary art is a relevant objective as well. In particular, the World Wide Web offers the opportunity of accessing large resources of low quality videos, images, and audios. Moreover, modern mobile devices allow people to record music and videos with inexpensive transducers that produce low quality data in respect to
	the expensive professional technology. Our restoration intent is therefore directly connected with the democratization of music production technologies and with the fruition of old audio recordings.}

The audio restoration literature is dominated by two general approaches: the first aims at reconstructing the sound as it was originally ``reproduced and heard by the people of the era'', while the second and most ambitious one aims at reconstructing ``the original sound source exactly as received by the transducing equipment (microphone, acoustic horn, etc.)''\cite{storm1980establishment,godsill1998digital}. However, an exact restoration is impossible in both cases. Particularly regarding the second approach, aiming at recovering the so-called ``true sound of the artist'' \cite{orcalli2001methodologies} exposes the restoration to subjective interpretations regarding the performer's artistic intention. Indeed, the artist's original intention is never completely captured by the recording because of the recording equipment limitations, such as microphones compression, noises, and degradation\cite{xu201905transferring}. To get over this issue, several studies tried to exploit the timbral features of the audio recording to compute original sound characteristics such as note intensities\cite{marinelli2020musical,weinzierl2018sound,jeong2017intensity}, but this is a particularly challenging problem hindered by the variability of MIDI velocity mappings\cite{dannenberg2006interpretation}.

Thus, we propose not to recover the original artistic intention but the intention survived until today and perceivable by the listener, as this is the best case scenario which is not influenced by subjective factors. The proposed idea consists in \begin{inparaenum}[a)] \item analyzing a recording via an AMT system so as to estimate the parameters of the performance and \item resynthesizing it using modern technologies. \end{inparaenum} In other words, we wish to retain the effort of the performer in a resynthesized version of the automatically-extracted music transcription.



Towards defining the specific problem, two concepts need to be distinguished, i.e. \textit{interpretation} and \textit{performance}. The \textit{performance} is the set of physical events that result in the activity of playing a music piece. It is bijectively associated with a certain time, place and performer, so that it is a unique unrepeatable act. \textit{Interpretation}, instead, refers to the \textit{ideal} performance that the performer has in mind and tries to realize. Thus, an interpretation could be repeated in different performances and differs from the performance because it lacks the adaptation to the context. It comprises the ultimate goal of the performer and thus, we identify it with the performer's artistic intention. During the restoration process, we seek to generate a new performance based on the interpretation extracted from the audio recording.

This operative definition is completely unrelated to the musicological debate about what an interpretation is -- e.g. when the notion ``interpretation'' was introduced with reference to music\cite{dreyfus2020beyond} or whether the interpretation is the ``performer's idea of the music''\cite{davies2001interpretation}. We rather focus on tracing the difference between  desired and realized performance, which differ due to external causes.

\added{

	\begin{figure}
		\centering
		\includegraphics[width=0.65\textwidth]{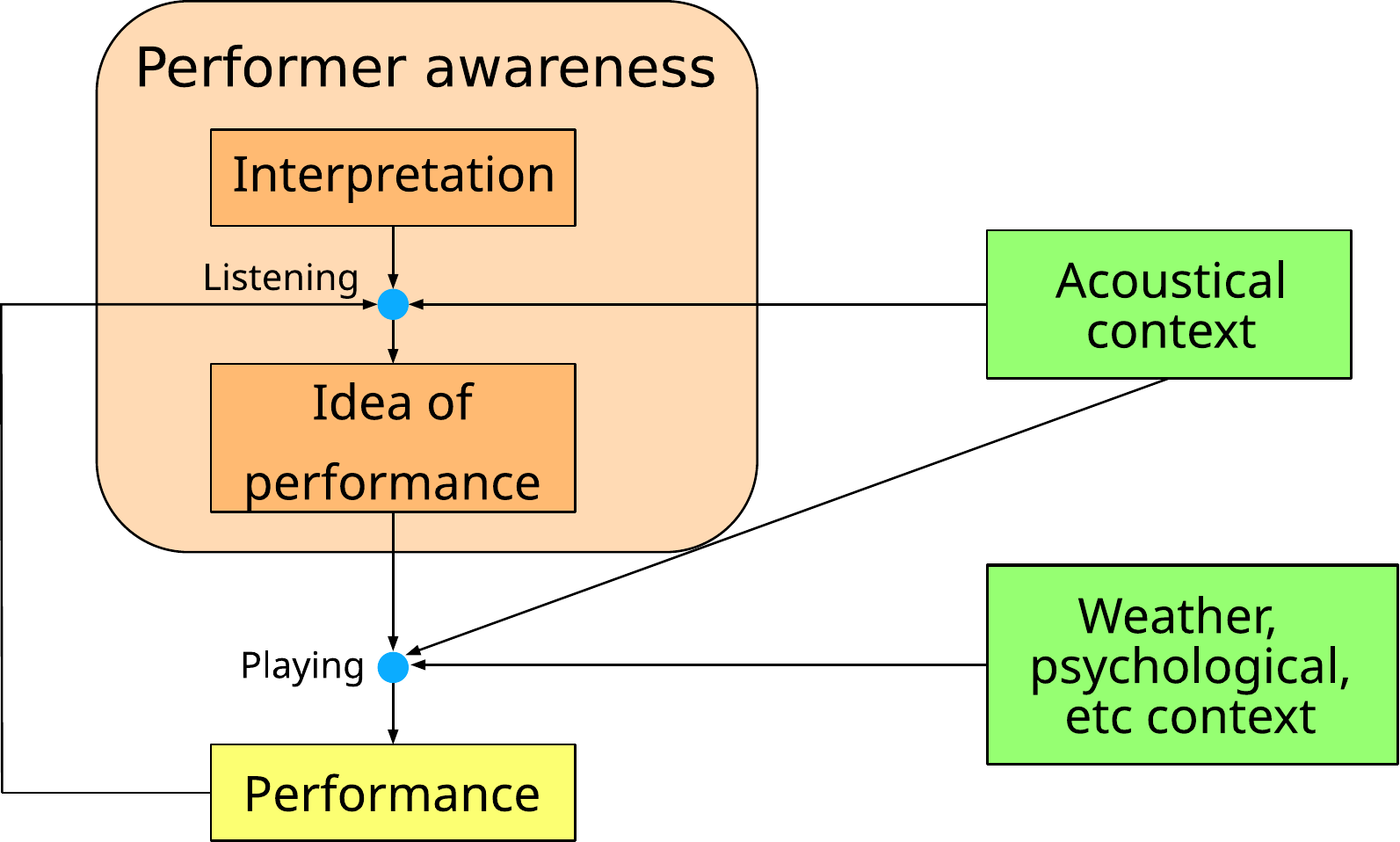}
		\caption{\added{Diagram showing the circular feedback involving the interpretation, the acoustical context, and the performance. The interpretation is influenced by the acoustical context based on the feedback coming from listening to the performance itself. The ``idea of performance'' is an intermediate representation that the performer consciously creates based on such feedback.}}
		\label{fig:diagram_interpretation}
	\end{figure}

	Such distinction is in line with the state-of-art research in the field of Music Performance Analysis that focuses on the acoustics of concert stages and rooms.
	The interest in the influence of the room
	acoustics on the performance dates back to 1968{~\cite{vonbekesy1968feedback}},
	but it has not received significant attention by the research community in the successive
	decades with the exception of a limited amount of works{~\cite{ternstrom1989longtime,
				naylor1992laboratory,bolzinger1994study}}. All these studies showed that
	musicians (orchestra, choir, piano, and percussion players) adapt their music
	performance to the acoustic environment in which they perform. Several music
	psychologists hypothesized the existence of an interior representation of the
	sound that the musician wants to convey{~\cite{gabrielsson1999performance}}. Such a
	perspective was further elaborated by various authors in an attempt to
	understand how musicians adapt their performance to various acoustic
	environments. First, subjective tests on musicians playing in different
	virtualized acoustic settings were explored and a circular feedback model between the
	performer and acoustical environment was
	proposed{~\cite{ueno2005experimental,ueno2005cognitive}}. Then, in the last
	decade, a few studies attempted to tackle the problem with objective
	evaluations. In 2010 and 2015, the same authors proposed two new studies in
	which physical features extracted from audio recordings were compared with the
	subjective self-evaluation of musicians and of listeners{~\cite{ueno2010effect,
				kato2015effect}}. From the comparison of objective and subjective evaluations,
	they argued that the feedback process was conscious. Other researchers tried to understand which factors of the room acoustics
	influence the performance and how~\cite{kalkandjiev2015influence}, arguing that
	the way in which musicians change their execution is performer-specific. In
	recent years, the research in the room influence on the performance has
	continued with the analysis of
	singers~\cite{luizard2018singers,potovcan2020estetska} and trumpet
	players~\cite{gari2019analysis}.  The overall contribution of the previous
	studies is that the adaptations applied by musicians influence the timbre, the
	amplitude dynamics, and the timing. An overview of existing works and
	methodologies has been recently published~\cite{kob2020effect}. However, all the
	existing studies, are directed towards the understanding of the factors characterizing  room acoustics. At the same time, they rarely consider the listener perception and never take
	into account indirect factors that can effectively change the acoustics of the
	instrument, such as the temperature and the humidity.

	Two theoretical concepts have been developed in the related literature, as outlined
	above. First, previous studies suggest the existence of a \textit{circular feedback} between the performer and the surrounding environment{~\cite[Fig. 1]{ueno2010effect}}. Second, the outlined literature is coherent with the existence of an \textit{interior representation} of the music performance that has to be realized; this idea was proposed by psychological studies{~\cite{gabrielsson1999performance}}, and developed in the above-mentioned literature{~\cite{kalkandjiev2015influenceb}}. The definition of ``interpretation'' proposed in this work is undoubtedly similar to the concept of such interior representation. In Figure{~\ref{fig:diagram_interpretation}}, the difference
	between ``interpretation'' and ``performance'' is clearly outlined.
}


\section{Designing the Test}\label{sec:design}

\begin{figure*}
	\centering
	\includegraphics[width=1\textwidth]{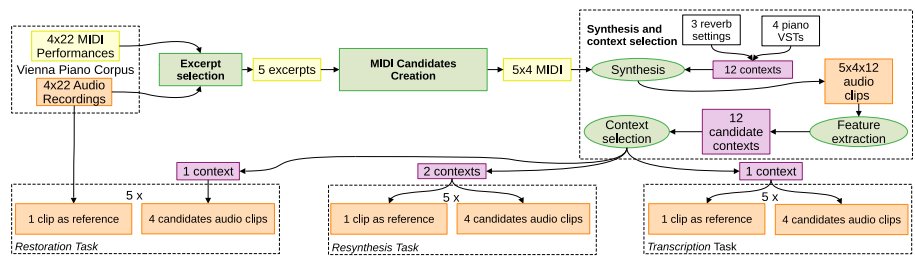}

	\caption{The workflow used for creating the restoration, resynthesis, and transcription tasks. Legend: a) Yellow: MIDI data; b) Orange:  audio data; c) Purple: contexts; d) Green: Operations.
	}
	\label{fig:scheme}
\end{figure*}
This section analyzes the conducted experiment from the technical point of view as well as the reasoning behind the presented choices.


\subsection{Research questions}

Given the definitions of Sec.~\ref{sec:concepts}, we assume the following:
\begin{enumerate}
	\item MIDI -- and consequently SMF -- is able to record every aspect characterizing a piano performance;
	\item using the same interpretation, a musician is going to create different performances given that the context (instrument, room, audience) changes;
	\item during the audio recording process, there is information loss up to a certain extent, while a different type of information, related to the context (including microphones),  may be introduced. Contextual alterations render practically impossible to extract the exact MIDI performance from the audio.
\end{enumerate}

The first research question that we seek to answer with the listening test is to which extent the interpretation is still identifiable when changing the context and retaining the performance, i.e. the MIDI recording. \added{According to our second assumption and to psychological studies described in Section~\ref{sec:concepts}, if the context changes and the performance stays constant, we assume that the two performances are generated from two different interpretations. Consequently, while retaining the same interpretation in two different contexts, we expect that the listener will perceive two different interpretations. With this research question,} we aim at assessing whether MIDI is effective in representing aspects related to interpretation.


The second research question is whether state-of-art AMT systems, typically trained to extract performance parameters, are effective in the extraction of the interpretation as well. Since it is practically impossible to recreate the exact performance, it is interesting to investigate whether a slightly modified MIDI coming from AMT systems is able to encompass interpretation aspects. In case this is true, we could resynthesize a given interpretation with a different context and obtain a different performance of the same interpretation.

Finally, following the line of thought presented in \cite{ycart2020investigating}, we want to provide a perceptual evaluation of AMT systems. 

\subsection{Tasks}\label{sec:task}

Three different tasks were designed addressing the above-mentioned questions. They consist in assessing the similarity in the interpretation between a reference audio excerpt and several candidates. The tasks differed in the way the audio clips were generated, as follows:

\begin{enumerate}
	\item in the first task, named ``transcription'', all audio excerpts including the reference were synthesized from MIDI using the same context (i.e., same virtual instrument and reverberation); 
	\item in the second task, named ``resynthesis'', all audio excerpts were still synthesized from MIDI but we used two different contexts for the reference and the candidates;
	\item in the third task, named ``restoration'', the reference was a real-world recording, while the candidates were synthesized from MIDI with a virtual instrument. Since the original recording contains substantially more noise than the synthesized candidates, this specific task is representative of a restoration process.
\end{enumerate}
All tasks used the same 5 excerpts extracted following the process explained in section \ref{sec:chosing}, where we also describe the reasoning behind the choice of the virtual instruments.

\subsection{Protocol and interface}\label{sec:interface}
Due to COVID-19 restrictions, we designed an online test using the ``Web Audio Evaluation Tool'' (WAET)\cite{jillings2015audio}.

First, subjects were prompted with some introductory slides explaining the difference between interpretation and performance. Specifically, after the formal definitions, they were suggested to adopt the following way of thinking: the interpretation is associated with the pianist, while the performance is related to a particular concert. Then, they listened to the first 30 seconds of two different performances by Maurizio Pollini of the Sonata No.30 op.109 by L. van Beethoven, and to a performance of the same piece by Emil Gilels; they were told that the first two were the same interpretation but different performances while the latter was both a different performance and interpretation. Finally, they were asked to
\begin{inparaenum}[a)]
	\item use headphones or headset,
	\item stay in a quiet place, and
	\item consent to the use of their answers in an anonymous form.
\end{inparaenum}

Next, subjects were asked preliminary questions, namely:
\begin{enumerate}
	\item what level of expertise they have with music (options: none/hobbyist vs. undergraduate/graduate/professional)
	\item how often they listen to classical music (options: less vs. more than 1 hour per week)
	\item how often they listen to music other than classical (options: less vs. more than 1 hour per week)
	\item what is the cost of the headphones they were using (options: less vs. more than 20 euros)
\end{enumerate}

\begin{figure*}
	\centering
	\includegraphics[width=\textwidth]{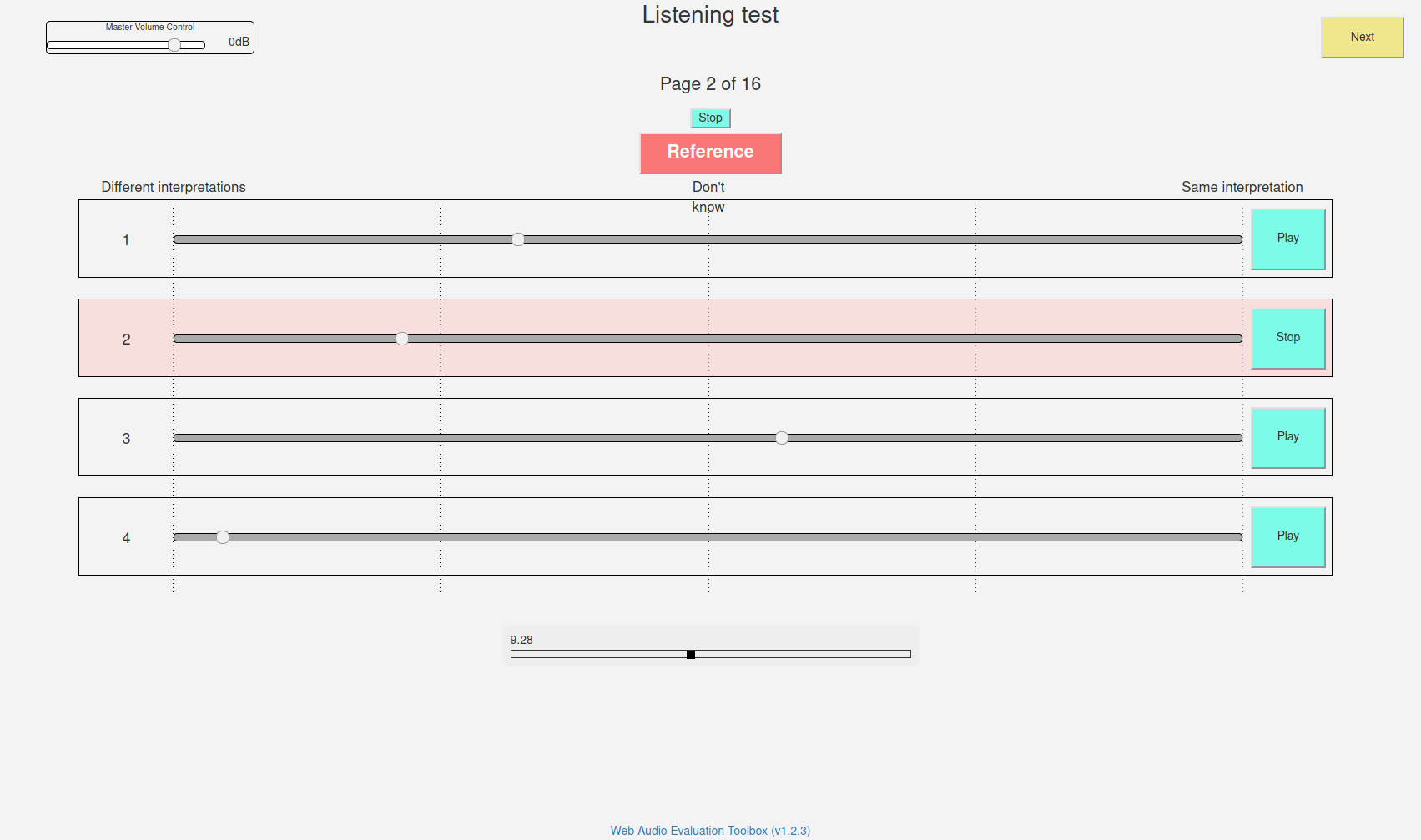}
	\caption{Screenshot of the interface created using the ``Web Audio Evaluation Tool''\cite{jillings2015audio}}
	\label{fig:screenshot}
\end{figure*}

At this point, subjects were exposed to the experimental interface through
an example question, with guitar instead of piano recordings.
Similarly to MUSHRA test\cite{feiten2005audio}, subjects could play back a reference audio file and four additional candidates containing the same musical excerpt resynthesized with various contexts and performances; they were asked to rate each candidate with a horizontal slider
on a continuous variable evaluating the extent to which the candidate clip contained the same interpretation as the reference. Three labels were put along the slider: ``different interpretations'', ``don't know'', and ``same interpretation'' -- see Fig. \ref{fig:screenshot} for a screenshot. \added{Users were instructed to experiment with the example question until s/he felt comfortable. As shown in Figure{~\ref{fig:screenshot}}, users could adjust the audio level at any time, even during the example question. Audio clips were normalized to -23 LUFS level in both the example and the actual test. However, authors are not aware of studies about loudness influence in Music Performance Assessment studies.}

Finally, subjects started the actual test, with piano recordings.
The order of questions and candidates were randomized, as well as the initial positions of the sliders to prevent biases. Since the test lasted approximately 30 minutes on average, if a subject decided to leave the session her/his answers were recorded. For this, we used a feature provided by WAET to first prompt new subjects with questions for which fewer answers were collected, so that the number of answers per question was uniform.

\subsection{Number vs. duration of excerpts}

In any listening test which deals with the artistic expression of the performer, an issue arises concerning the length of the excerpts. In general, it can be expected that longer duration will lead to more accurate subjective judgments. However, one second competing factor is the total duration of the test: the longer the test is, the more difficult is to find volunteers willing to take the full test and be able to keep their concentration for the entire duration\cite{schwarz2016effects}.

One study observed that the duration of excerpts did not influence the emotional response of the listeners\cite{bigand2005multidimensional}. On the other hand, it was shown that in the context of piano performance assessment, graduate music students and faculty professors rated $60$~s excerpts higher than $20$~s excerpts, while no significant difference was observed for non-graduate students\cite{wapnick2005effects}. Another study conducted on wind bands took into account the level of the performers and showed that music majors rated $25$~s and $50$~s excerpts higher than $12$~s excerpts in the case of university or professional level performances, while the opposite happened for performances at the high-school level\cite{geringer2007effects}. A study on children chorale music revealed that $60$~s excerpts were rated higher then $20$~s excerpts by music majors\cite{napoles2009effect}. Finally, another study on band performances was conducted with excerpts $12$~s, $15$~s, and $30$~s long, and found that ratings from music majors were higher for long excerpts of bad performances\cite{williams2016effect}.

It comes from the cited literature, that a minimum sufficient duration to observe the difference between ratings given by expert and non-expert listeners is in the range $15-25$~s. Thus, we used excerpts lasting $20$~s. Since each subject performed all of the three tasks -- see Sec.~\ref{sec:task} -- and we aimed at keeping the test less than 30 minutes long, we opted for 5 excerpts, resulting in 15 questions (5 questions per task) each with 5 audio clips (1 reference and 4 candidates) lasting $20$~s, for a total minimum duration of 25 minutes.

The studies discussed above also suggest that expert listeners tend to base their judgements on longer time features with respect to non expert raters. We expect a similar behavior for the task of comparing two interpretations. As a consequence, if no significant difference is found between expert and non-expert ratings, a difference may still be observed when using excerpts longer than $20$~s. The only way to rule out such hypothesis would be by using full song excerpts.

\section{Generating excerpts and contexts}\label{sec:chosing}

Since we used a limited number of excerpts, we wanted to chose them in a way that minimizes any type of subjective bias. The same also applies for the choice of the contexts. In Fig. \ref{fig:scheme}, we show the overall workflow used for solve this problem.

\subsection{The $p$-dispersion problem and uniform selection}

\begin{figure}
	\centering
	\includegraphics[width=0.5\textwidth]{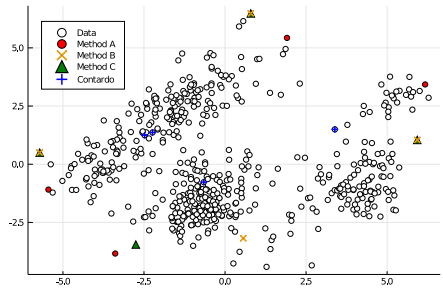}

	\caption{Comparison of methods for the $p$-dispersion problem with $p=4$. Data are the windows extracted from the \textit{Vienna 4x22 Piano Corpus}. PCA was used in this plot to reduce the dimensionality from 15 down to 2 for demonstration purposes. For the listening test, we used \textit{Method A}. \textit{Contardo} is the state-of-art mathematically-proven method for the $p$-dispersion problem\cite{contardo2020decremental}. For the comparison, we used the original Julia code provided by the author.}
	\label{fig:p_dispersion}
\end{figure}

For choosing the excerpts, we built a dataset where each sample was represented by features extracted from audio clips and MIDI symbols. Since we expected that the perception of music changes as these features vary, we aim at maximizing the distance between feature vectors of the chosen excerpts so that perceptual variations are revealed.

This means that we are looking for the $p$ samples in the feature space which are distributed uniformly, while the distance between them is maximal.
This problem can be seen as a variation of the \textit{p-dispersion problem}\cite{erkut1990discrete} or \textit{max-min facility dispersion problem}\cite{ravi1994heuristic}. However, the $p$-dispersion problem does not impose any restriction with respect to data distribution, thus we derived four \textit{ad hoc} algorithms solving the present problem based on Ward-linkage clustering\cite{everitt2011hierarchical}.
Fig. \ref{fig:p_dispersion} compares them on our excerpt dataset.

\subsection{$p$-dispersion problem}

Here we briefly describe the algorithm and compare it with a state-of-art method for solving the $p$-dispersion problem but we leave to future works the mathematical study of the method\footnote{Code available at \url{https://framagit.org/sapo/selection_test}}.

Our approach consists in finding $p$ subsets using hierarchical clustering. We used Ward method because it tries to minimize the variance inside each cluster and, consequently, each cluster is well represented by its centroid. In contrast to $k$-means clustering, instead, it is well suitable even for little sized datasets and does not depends on initialization heuristic. \added{After having partitioned the data in $p$ clusters with the Ward method, we chose one point per cluster as follows.} We chose the point in each cluster which maximized the distance from the centroid of all the other points in the dataset (\textit{Method A}). Successively, we have also considered other strategies to chose the point in each cluster, namely: the point which maximize the minimum distance from the centroids of the other clusters (\textit{Method B}), the point which maximize the minimum distance from the points in the other clusters (\textit{Method C}), the point which maximize minimum distance from all the other points (\textit{Method D}). In table \ref{tab:pdispersion} we compare these methods to the state-of-art method for the $p$-dispersion problem by Contardo, for $p=[4, 5, 10]$, using the code provided us by the author; we used the datasets of the Contardo's work containing less than 10.000 samples with the addition of our dataset created with PCA output of 2 features per sample\cite{contardo2020decremental}. For method\cite{contardo2020decremental}, we used the Julia implementation provided us by the author. We have always used the euclidean distance -- or sometime the sum of the absolute differences for improving the computation time.

\begin{table}
	\center
	\caption{Comparison of methods for solving the $p$-dispersion problem. Columns are: average minimum distance in the output set, average time in seconds needed, percentage of instances in which each method had \textit{Min Dist} grater or equal than any other (all) or than the Contardo's method\cite{contardo2020decremental}}
	\label{tab:pdispersion}

	\begin{tabular}{r|c|c|c|c|}
		\cline{2-5}
		\multicolumn{1}{l|}{}                                  & \multicolumn{1}{l|}{\cellcolor[HTML]{C0C0C0}Min Dist} & \multicolumn{1}{l|}{\cellcolor[HTML]{C0C0C0}Time (s)} & \multicolumn{1}{l|}{\cellcolor[HTML]{C0C0C0}\% wins vs all} & \multicolumn{1}{l|}{\cellcolor[HTML]{C0C0C0}\% wins vs\cite{contardo2020decremental}} \\ \hline
		\multicolumn{1}{|r|}{\cellcolor[HTML]{C0C0C0}Method A} & 1.58E+04                                              & 2.00                                                  & 9.52\%                                                      & 60.32\%                                                                               \\ \hline
		\rowcolor[HTML]{EFEFEF}
		\multicolumn{1}{|r|}{\cellcolor[HTML]{C0C0C0}Method B} & 1.33E+04                                              & \textbf{1.66}                                         & 25.40\%                                                     & 61.90\%                                                                               \\ \hline
		\multicolumn{1}{|r|}{\cellcolor[HTML]{C0C0C0}Method C} & \textbf{1.90E+04}                                     & 9.34                                                  & \textbf{41.27\%}                                            & \textbf{63.49\%}                                                                      \\ \hline
		\rowcolor[HTML]{EFEFEF}
		\multicolumn{1}{|r|}{\cellcolor[HTML]{C0C0C0}Method D} & 8.43E+03                                              & 10.44                                                 & 1.59\%                                                      & 50.79\%                                                                               \\ \hline
		\multicolumn{1}{|r|}{\cellcolor[HTML]{C0C0C0}Contardo} & 9.21E+03                                              & 153.30                                                & 34.92\%                                                     & 100.00\%                                                                              \\ \hline
	\end{tabular}\end{table}


\subsection{Excerpt selection}\label{sec:selection}
The selection of the audio excerpts started from the \textit{Vienna 4x22 Piano Corpus}\cite{goebl1999vienna}, which consists of 88 audio and corresponding MIDIs recordings of 4 famous pieces highly representative of the classical-romantic music period, played by 22 professional and advanced student pianists. This corpus was useful in order to have a negative reference (NR) available for any chosen excerpt -- i.e. a different interpretation. We used the \textit{ASMD} framework\cite{simonetta2020automatic} to handle the loading of files and dispatching parallel processing routines. Every audio clip was converted to monophonic, downsampled to $22050$~Hz and normalized using ReplayGain\cite{replaygain}. MIDI files were loaded using pianorolls where each pixel contained the velocity value of the ongoing note and each column had a resolution of $5$~ms. In order to compensate for temporal misalignments, for each pair (\textit{audio, MIDI}), we identified the audio frame of first onset and last offset using an energy threshold of -60 \textit{dB} under which the sound was identified as silence, and trimmed the files accordingly.


We split the audio recordings in windows and considered each window as a possible excerpt for the listening test. Each window lasted exactly $20$~s with a hop-size of $10$~s, resulting in 564 total windows. For each window we extracted \added{a set of audio and symbolic features. The first were extracted from the audio recordings and consisted of high-level features among the most used in the MIR field, extracted using the state-of-the-art library Essentia{~\cite{bogdanov2013essentia}}.To take into account the timbral characteristics, we extracted 13 MFCCs{~\cite{alias2016review}}; we used a state-of-the-art onset detection method to extract} 7 rhythmic descriptors~\cite{zapata2014multifeature}. Furthermore, we used the Essentia library to estimate BPM, along with the first and second peak values, spreads, and weights of the corresponding histogram; \added{such features are related to the timing characteristics of the performance as rendered in the audio. Regarding the symbolic features, we used the non-zero pixels in the window pianorolls to extract information about the performance as recorded by the sensorized piano. Specifically,} we extracted the average and standard deviation of pitches, velocities, duration, number of contemporaneous notes in each column and pitch difference in each column relatively to the lowest pitch in the column -- which relates to the type of harmony. The resulting features were concatenated in an array of 30 features. Then, the 564 windows were standardized (mean removal and variance scaling) and passed through PCA to obtain 15 relevant features explaining 92\% of variance.

We applied the methods described earlier to look for 4 dispersed windows. To ensure that the 4 selected points well represented the whole dataset, we chose the only method that managed to select one window for each of the 4 pieces in the \textit{Vienna} corpus (\textit{Method A}). Then, we added one excerpt computed as the medoid of the dataset, obtaining in total 5 excerpts.

\begin{figure*}[t]
	\centering
	\includegraphics[width=\textwidth]{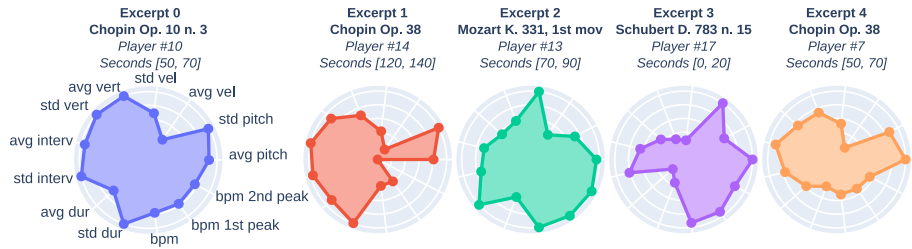}
	\caption{Features extracted from the chosen excerpts; they are normalized in [0,1]. Seconds are relative to the audio recordings.}
	\label{fig:excerpts}
\end{figure*}

It should be noted all the excerpts last exactly $20$~s; this may be criticized as it implies that NR and realignments could lead to slightly different contents, since a given excerpt could be played and transcribed in a different time lapse. On the other hand, having the NR lasting a different amount of time would have produced a potential bias at the listener side, who would have been able to identify the NR based on duration rather than audio content. The variability of the features of the chosen excerpts are shown in Fig. \ref{fig:excerpts}.

\subsection{MIDI Candidates Creation}\label{sec:models}

For each  question in the test, we used four candidates in addition to the reference excerpt: a negative reference (NR) containing a performance by another pianist of the same excerpt, a hidden reference (HR) containing the same performance as the reference, and two performances extracted using two different AMT systems described next. For the NR, we realigned its MIDI recording to the chosen reference MIDI using FastDTW\cite{salvador2007toward}, before trimming. For the \textit{restoration} task, where the reference was a real-world recording, the HR was not the same audio recording (which would be immediately recognizable from the remaining candidates), but rather the associated MIDI available in the Vienna corpus.

Various AMT models were published in recent years; unfortunately, only the ones trained for solo piano music achieve satisfactory performances. In particular one model, here called \textit{onsets and frames} (O\&F), has been extensively evaluated on various datasets and has been shown to overcome the rest in almost every piece\cite{hawthorne2018onsets,ycart2020investigating}. Recently, it has been shown that AMT models could be enhanced by pre-stacking a U-Net\cite{wu2019polyphonic}. U-Nets were first used for image segmentation and then for audio source separation. By pre-stacking a U-net, the network tends to learn knowledge regarding the sparse structure of the spectrogram-like input representation. However, we are more interested in understanding how audio-based AMT differs from score-informed AMT; intuitively, since score-informed AMT models exploit more information, the output should be more accurate. Thus, we compared O\&F with a score-informed model (SI) which we developed based on a previous work\cite{jeong2020noteintensity}.


\added{In SI, inputs are a non-aligned score and an audio recording, while the output is a list of MIDI notes, each associated with onset, offset and velocity. SI performs audio-to-score alignment using a method based on Dynamic Time Warping that improves a previous system for piano music\cite{kwon2017audiotoscore}, and subsequently executes a Non-negative Matrix Factorization as source-separation method for each piano note. Then, it employs a neural network for estimating the velocities of each aligned note using as input the spectrogram of the note, computed thanks to the source-separation. Since the SI method requires the score, one was obtained from the World Wide Web for each of the 5 excerpts. For further details see the \textit{Supplementary Materials}.
}

Both the AMT systems predict pitches, onsets, offsets and velocities, while no other MIDI parameter (e.g., pedaling, etc.) is considered.

\subsection{Synthesis and Context selection}
After producing the MIDI files, we synthesized them using 4 high quality different virtual pianos:
\begin{inparaenum}[a)]
	\item the free \textit{Salamander Grand Piano},\footnote{free as in speech: https://musical-artifacts.com/artifacts/533}
	\item two Pianoteq instruments freely available for research purposes (\textit{Grand Steinway B} and \textit{Grand Grotrian}), and
	\item the \textit{Steinway} piano from Garritan Personal Orchestra 4.
\end{inparaenum}
We post-processed every synthesized MIDI using SoX\footnote{sox.sourceforge.net} in order to add reverberation using  two different settings (values $50$ and $100$ of the SoX's \texttt{reverb} option). Thus, $12$ different contexts, i.e. $4$ without and $8$ with reverberation, were formed.

We synthesized all the MIDI files with each context obtaining $12$ different sets of audio clips. We extracted $13$ MFCC and $7$ rhythmic descriptors from each audio clip and computed the mean features to represent each context. We chose the medoid context for the \textit{transcription} task and the most distant context from the average features of the original audio recordings for the \textit{restoration} task. For the \textit{resynthesis} task, two contexts were needed: in this case PCA explaining 99\% of variance was applied to obtain a 10 dimensional representation from the original 20, and then searched for the two farthest points in the feature space based on the euclidean distance.

As a result of this process, the selected contexts were:
\begin{enumerate}[a)]

	\item the Pianoteq \textit{Grand Steinway} with SoX reverb set to 100 for the transcription task \added{(44.1KHz/16bit stereo audio for both reference and candidates)};

	\item the Pianoteq \textit{Grand Steinway} without reverb and the \textit{Salamander Grand Piano} with SoX reverb set to 50 for the resynthesis task, candidates and reference respectively \added{(44.1KHz/16bit stereo for the reference and 44.1KHz/24bit stereo for candidates)};

	\item the \textit{Salamander Grand Piano} without reverb for the restoration task \added{(44.1KHz/16bit mono for the reference and 44.1KHz/24bit stereo for candidates)}.
\end{enumerate}

\added{Note that no perceptual difference is known between 24 and 16 bit depth at the same sample rate~\cite{repp2006recording, mortberg2007dithered,mizumachi2017discussion}.}

\section{Results}\label{sec:results}

\begin{figure*}
	\centering
	\includegraphics[width=0.65\textwidth]{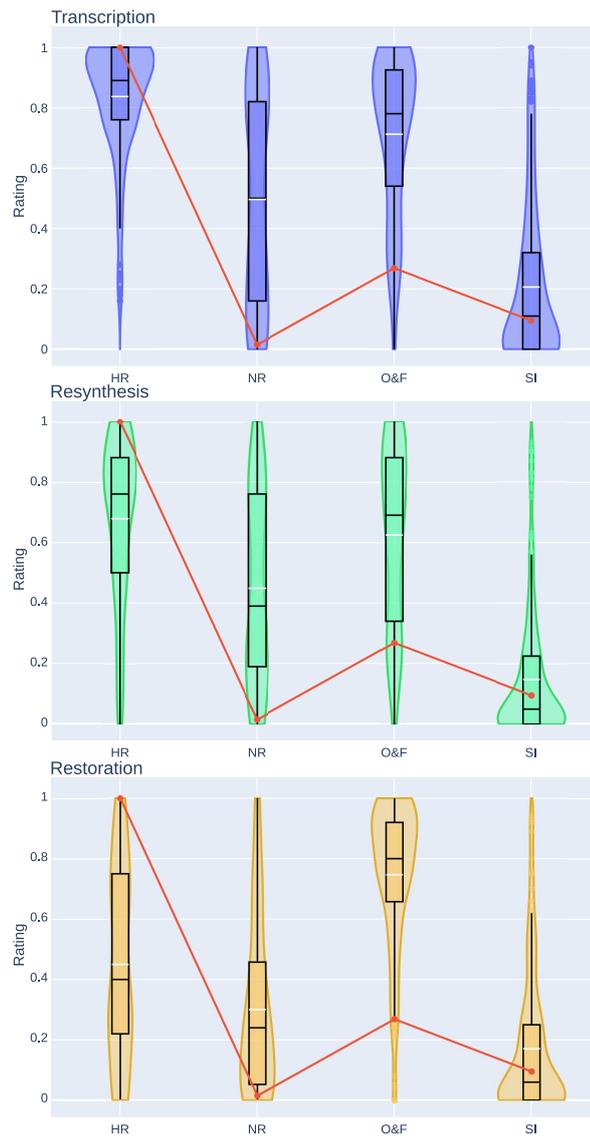}
	\caption{Ratings per task averaged over all the excerpts. The red line identical in all tasks is the objective F-measure. White horizontal line is the mean, the black horizontal line is the median. All plotted distributions pass the pairwise significance tests against the other distributions in the same task, except for O\&F and HR in the resynthesis task. \added{Plots are created with{~\cite{plotly}}.}}
	\label{fig:average}
\end{figure*}

\begin{table}
	\centering
	\caption{Number of questions with $p>0.05$ for each pair of methods \added{for both Sudents' t and Wilcoxon tests}.}

	\label{tab:pvalues}
	\begin{tabular}{cccc}
		\cline{2-2}
		\multicolumn{1}{c|}{SI}   & \multicolumn{1}{c|}{4} &                        &                        \\ \cline{2-3}
		\multicolumn{1}{c|}{O\&F} & \multicolumn{1}{c|}{6} & \multicolumn{1}{c|}{0} &                        \\ \cline{2-4}
		\multicolumn{1}{c|}{HR}   & \multicolumn{1}{c|}{7} & \multicolumn{1}{c|}{1} & \multicolumn{1}{c|}{9} \\ \cline{2-4}
		                          & NR                     & SI                     & O\&F
	\end{tabular}
\end{table}
The listening test was communicated through mailing lists, chats, university courses, etc.; in total, 91 subjects responded to the entire test. Thanks to JavaScript-based WAET, we observed the subjects' behavior during the test, so that we were able to discard the answers where subjects listened to the excerpt for less than 5 seconds or where they did not move the cursor. After such filtering, we obtained more than 40 answers per question. Since these did not result in enough answers for each class of the initial questionnaires described in \ref{sec:interface}, we focused on two groupings only:

\begin{itemize}
	\item subjects listening to classical music less than 1 hour per week (50) vs. subjects listening to classical music more often (41);
	\item subjects who have never studied music/hobbyists (57) vs. subjects who studied music professionally or having a degree or working as musicians (34);
\end{itemize}

We observed a general trend of non-experts providing higher ratings, meaning that, with respect to more experienced listeners or musicians, they rated candidates to be more like the same interpretation as the reference. However, this difference was not always statistically significant, thus not useful for the sake of our research questions. According to the literature discussed in Sec.~\ref{sec:design}, we could expect that by using longer excerpts the difference in the ratings would become significant and that expert listeners could give more accurate ratings.

\added{We collected an imbalanced number of answers per type of headphones, namely 22 for headphones costing less than 20 euros and 69 for headphones costing more than 20 euros. Since we have found contrasting studies in literature about possible correlations between headphone retailing cost and sound quality, we have decided to disregard this factor during the successive analysis~\cite{breebaart2017correlation,gutierrezparera2018perception}.}

During control group based analysis, we had more than 20 answers available per question and control group. \added{We first applied Shapiro-Wilk normality tests with Bonferroni-Holm correction and $\alpha = 0.05$ to the collected responses for each control group, question, and method. We observed that the null-hypothesis -- i.e. the collected answers are normally distributed -- was rejected depending on the method and on the question. Consequently, we have performed the whole statistical analysis with both parametric and non-parametric methods and leave to the reader the ability to decide which test should be taken into account based on the single case. The following discussion and conclusions, however, hold in both the two cases -- i.e. parametric and non-parametric analysis.}

We computed error margins at 95\% of confidence\added{, that is a quantification of the accuracy for the estimated mean $\hat{\mu}$: we can say that by resampling the distribution, 95\% of the collected populations will have the mean in $\hat{\mu} \pm e$, where $e$ is the error margin. Error margins were computed with both normal distribution assumption~\cite{tanur2011margin} -- i.e. parametric estimation -- and bootstrapping methods~\cite{chernick2011bootstrap} -- i.e. non-parametric estimation. The full set of error margins is available in the \textit{Supplementary Materials}. For our discussion, we can say that parametric and non-parametric error margins were rarely different when rounded at the 2nd decimal digit and that they} ranged between 2\% and 17\%. Without using control groups, the error margins were between 3\% and 10\%, and between 2\% and 5\% when we average the ratings over the questions of the same task.

We analyzed results using \added{ANOVA} and Kruskal-Wallis tests with $\alpha=\added{0.01}$ and we rejected the null hypothesis in all considered questions. We further analyzed the data using the Wilcoxon \added{and the Student's t-test for related variables} with the Bonferroni-Holm correction and $\alpha=0.05$. In case the test condition is not satisfied, we cannot reject the hypothesis according to which the perception of two candidates is explainable by the same model. This is usually observable for non-expert subjects. Table~\ref{tab:pvalues} shows a general overview of the $p$-values computed for each pair of methods.

Fig. \ref{fig:average} illustrates the average ratings for every task. The HR and the NR were recognized in all tasks, and O\&F performed always better than NR and SI.
In a post-experimental interview, several subjects reported that SI was hardly comparable to the reference due to bad alignment. Indeed, in some excerpts, notes were distributed by SI in a very short time, producing a correspondingly long-lasting silence; this was often caused by missing/extra notes appearing in the music score. Thus, we conclude that the main reason for which SI was always rated worse than O\&F is related to misalignments; we can consequently answer the third research question by stating that score-informed approaches are generally limited by the alignment stage and that, as of now, monomodal AMT approaches provide improved performance assessed from a perceptual point of view.

\begin{figure*}
	\centering
	\includegraphics[width=0.65\textwidth]{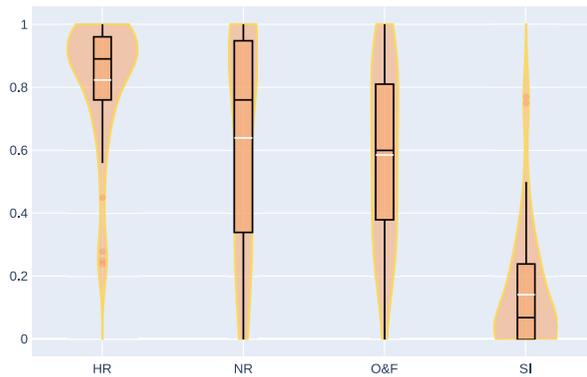}
	\caption{Ratings for Excerpt 3 in the Transcription task. For this excerpt, all distributions pass the pairwise significance test except O\&F and NR. \added{Plots are created with{~\cite{plotly}}.}}
	\label{fig:transcr3}
\end{figure*}

In the \textit{restoration} task, O\&F was rated higher than HR.
Since this behavior is not observable in the \textit{transcription} and \textit{resynthesis} tasks, and since the MIDI files were identical throughout all tasks, we attribute this outcome to the specificity of the \textit{restoration} task, where the HRs (MIDI) were different from the references (audio), unlike the other tasks. In particular, at the time of writing, performance annotations in the \textit{ASMD} framework do not include information about the pedaling used by the players, and recorded in the audio. Thus the HRs in the \textit{restoration} task were synthesized from MIDI with no sustain control changes, whereas the audio references contained them. On the other hand, O\&F is not able to transcribe the pedaling, but its authors enlarged the duration of sustained notes in the training ground-truth, so that the prediction of note duration is temporally tied to the duration of the resonance of the note rather than the onset/offset of the key. Such a durational enlargement allows O\&F (and SI as well, as it uses O\&F for the alignment) to predict duration perceptually more accurate than the HR in the \textit{restoration} task. 

The \textit{resynthesis} task is also worth some further discussion. In this task, HR and O\&F are perceived similarly, especially by non-expert listeners, and there are no statistically significant differences between the distributions of their ratings -- see Fig.~\ref{fig:average}. Even though HR is rated slightly higher than O\&F, it can be noted to score lower than in the \textit{transcription} task. This suggests that the whole reference interpretation was hardly recognised in the HR, and that part of the interpretation was perceptually lost.
Based on this outcome, we can try to address negatively the first research question, that is: when the context changes, MIDI representation seems not adequate to reproduce the same interpretation. However, other experiments are needed to confirm this hypothesis.

Analysing results question by question, we discovered an interesting behavior in the \textit{transcription} task for excerpt 3 -- see Fig. \ref{fig:excerpts} and \ref{fig:transcr3}. There, O\&F is associated with lower ratings than NR with $p=0.38$ \added{for Wilcoxon test and $p=0.47$ for Student's t-test}, meaning that the transcription is so inadequate that another interpretation resembles better the original one. Similar results are derived when investigating excerpt 0
in \textit{resynthesis} and \textit{transcription} task, where NR and O\&F present almost identical ratings ($p>0.23$). This behavior is more evident when looking at less expert listeners. Such results can also answer negatively to the second research question: the state of the art for piano AMT may not be able to extract parameters usable for reproducing the original interpretation, regardless changes in context (resynthesis and transcription tasks).

\section{A new measure}\label{sec:measure}

\begin{table}
	\center
	\caption{Correlations of various measures with the average rating of the subjects. Values are percentages.}
	\label{tab:correlations}
	\begin{tabular}{ccc|c|c|c|}
		\cline{4-6}
		\rowcolor[HTML]{FFFFFF}
		                                                                                                                                            &                                                                                  &              & \cellcolor[HTML]{C0C0C0}\textbf{OBJ} & \cellcolor[HTML]{C0C0C0}\textbf{PEAMT} & \cellcolor[HTML]{C0C0C0}\textbf{Ours} \\ \hline
		\rowcolor[HTML]{C0C0C0}
		\multicolumn{1}{|c|}{\cellcolor[HTML]{C0C0C0}}                                                                                              & \multicolumn{1}{c|}{\cellcolor[HTML]{C0C0C0}}                                    & \textbf{min} & \cellcolor[HTML]{EFEFEF}49           & \cellcolor[HTML]{EFEFEF}62             & \cellcolor[HTML]{EFEFEF}\textbf{95}   \\ \cline{3-6}
		\rowcolor[HTML]{C0C0C0}
		\multicolumn{1}{|c|}{\cellcolor[HTML]{C0C0C0}}                                                                                              & \multicolumn{1}{c|}{\cellcolor[HTML]{C0C0C0}}                                    & \textbf{max} & \cellcolor[HTML]{FFFFFF}78           & \cellcolor[HTML]{FFFFFF}\textbf{99}    & \cellcolor[HTML]{FFFFFF}97            \\ \cline{3-6}
		\rowcolor[HTML]{C0C0C0}
		\multicolumn{1}{|c|}{\cellcolor[HTML]{C0C0C0}}                                                                                              & \multicolumn{1}{c|}{\multirow{-3}{*}{\cellcolor[HTML]{C0C0C0}\textbf{Pearson}}}  & \textbf{avg} & \cellcolor[HTML]{FFFFFF}75           & \cellcolor[HTML]{FFFFFF}89             & \cellcolor[HTML]{FFFFFF}\textbf{97}   \\ \cline{2-6}
		\rowcolor[HTML]{C0C0C0}
		\multicolumn{1}{|c|}{\cellcolor[HTML]{C0C0C0}}                                                                                              & \multicolumn{1}{c|}{\cellcolor[HTML]{C0C0C0}}                                    & \textbf{min} & \cellcolor[HTML]{EFEFEF}40           & \cellcolor[HTML]{EFEFEF}40             & \cellcolor[HTML]{EFEFEF}\textbf{80}   \\ \cline{3-6}
		\rowcolor[HTML]{C0C0C0}
		\multicolumn{1}{|c|}{\cellcolor[HTML]{C0C0C0}}                                                                                              & \multicolumn{1}{c|}{\cellcolor[HTML]{C0C0C0}}                                    & \textbf{max} & \cellcolor[HTML]{FFFFFF}\textbf{100} & \cellcolor[HTML]{FFFFFF}80             & \cellcolor[HTML]{FFFFFF}80            \\ \cline{3-6}
		\rowcolor[HTML]{C0C0C0}
		\multicolumn{1}{|c|}{\multirow{-6}{*}{\cellcolor[HTML]{C0C0C0}\textbf{Transcription}}}                                                      & \multicolumn{1}{c|}{\multirow{-3}{*}{\cellcolor[HTML]{C0C0C0}\textbf{Spearman}}} & \textbf{avg} & \cellcolor[HTML]{FFFFFF}\textbf{80}  & \cellcolor[HTML]{FFFFFF}\textbf{80}    & \cellcolor[HTML]{FFFFFF}\textbf{80}   \\ \hline
		\rowcolor[HTML]{C0C0C0}
		\multicolumn{1}{|c|}{\cellcolor[HTML]{C0C0C0}}                                                                                              & \multicolumn{1}{c|}{\cellcolor[HTML]{C0C0C0}}                                    & \textbf{min} & \cellcolor[HTML]{EFEFEF}41           & \cellcolor[HTML]{EFEFEF}64             & \cellcolor[HTML]{EFEFEF}\textbf{95}   \\ \cline{3-6}
		\rowcolor[HTML]{C0C0C0}
		\multicolumn{1}{|c|}{\cellcolor[HTML]{C0C0C0}}                                                                                              & \multicolumn{1}{c|}{\cellcolor[HTML]{C0C0C0}}                                    & \textbf{max} & \cellcolor[HTML]{FFFFFF}73           & \cellcolor[HTML]{FFFFFF}\textbf{100}   & \cellcolor[HTML]{FFFFFF}99            \\ \cline{3-6}
		\rowcolor[HTML]{C0C0C0}
		\multicolumn{1}{|c|}{\cellcolor[HTML]{C0C0C0}}                                                                                              & \multicolumn{1}{c|}{\multirow{-3}{*}{\cellcolor[HTML]{C0C0C0}\textbf{Pearson}}}  & \textbf{avg} & \cellcolor[HTML]{FFFFFF}66           & \cellcolor[HTML]{FFFFFF}85             & \cellcolor[HTML]{FFFFFF}\textbf{98}   \\ \cline{2-6}
		\rowcolor[HTML]{C0C0C0}
		\multicolumn{1}{|c|}{\cellcolor[HTML]{C0C0C0}}                                                                                              & \multicolumn{1}{c|}{\cellcolor[HTML]{C0C0C0}}                                    & \textbf{min} & \cellcolor[HTML]{EFEFEF}40           & \cellcolor[HTML]{EFEFEF}\textbf{60}    & \cellcolor[HTML]{EFEFEF}\textbf{60}   \\ \cline{3-6}
		\rowcolor[HTML]{C0C0C0}
		\multicolumn{1}{|c|}{\cellcolor[HTML]{C0C0C0}}                                                                                              & \multicolumn{1}{c|}{\cellcolor[HTML]{C0C0C0}}                                    & \textbf{max} & \cellcolor[HTML]{FFFFFF}\textbf{100} & \cellcolor[HTML]{FFFFFF}\textbf{100}   & \cellcolor[HTML]{FFFFFF}80            \\ \cline{3-6}
		\rowcolor[HTML]{C0C0C0}
		\multicolumn{1}{|c|}{\multirow{-6}{*}{\cellcolor[HTML]{C0C0C0}\textbf{Resynthesis}}}                                                        & \multicolumn{1}{c|}{\multirow{-3}{*}{\cellcolor[HTML]{C0C0C0}\textbf{Spearman}}} & \textbf{avg} & \cellcolor[HTML]{FFFFFF}\textbf{80}  & \cellcolor[HTML]{FFFFFF}\textbf{80}    & \cellcolor[HTML]{FFFFFF}\textbf{80}   \\ \hline
		\rowcolor[HTML]{C0C0C0}
		\multicolumn{1}{|c|}{\cellcolor[HTML]{C0C0C0}}                                                                                              & \multicolumn{1}{c|}{\cellcolor[HTML]{C0C0C0}}                                    & \textbf{min} & \cellcolor[HTML]{EFEFEF}-4           & \cellcolor[HTML]{EFEFEF}55             & \cellcolor[HTML]{EFEFEF}\textbf{78}   \\ \cline{3-6}
		\rowcolor[HTML]{C0C0C0}
		\multicolumn{1}{|c|}{\cellcolor[HTML]{C0C0C0}}                                                                                              & \multicolumn{1}{c|}{\cellcolor[HTML]{C0C0C0}}                                    & \textbf{max} & \cellcolor[HTML]{FFFFFF}57           & \cellcolor[HTML]{FFFFFF}94             & \cellcolor[HTML]{FFFFFF}\textbf{100}  \\ \cline{3-6}
		\rowcolor[HTML]{C0C0C0}
		\multicolumn{1}{|c|}{\cellcolor[HTML]{C0C0C0}}                                                                                              & \multicolumn{1}{c|}{\multirow{-3}{*}{\cellcolor[HTML]{C0C0C0}\textbf{Pearson}}}  & \textbf{avg} & \cellcolor[HTML]{FFFFFF}29           & \cellcolor[HTML]{FFFFFF}78             & \cellcolor[HTML]{FFFFFF}\textbf{89}   \\ \cline{2-6}
		\rowcolor[HTML]{C0C0C0}
		\multicolumn{1}{|c|}{\cellcolor[HTML]{C0C0C0}}                                                                                              & \multicolumn{1}{c|}{\cellcolor[HTML]{C0C0C0}}                                    & \textbf{min} & \cellcolor[HTML]{EFEFEF}0            & \cellcolor[HTML]{EFEFEF}0              & \cellcolor[HTML]{EFEFEF}\textbf{80}   \\ \cline{3-6}
		\rowcolor[HTML]{C0C0C0}
		\multicolumn{1}{|c|}{\cellcolor[HTML]{C0C0C0}}                                                                                              & \multicolumn{1}{c|}{\cellcolor[HTML]{C0C0C0}}                                    & \textbf{max} & \cellcolor[HTML]{FFFFFF}60           & \cellcolor[HTML]{FFFFFF}60             & \cellcolor[HTML]{FFFFFF}\textbf{100}  \\ \cline{3-6}
		\rowcolor[HTML]{C0C0C0}
		\multicolumn{1}{|c|}{\multirow{-6}{*}{\cellcolor[HTML]{C0C0C0}\textbf{Restoration}}}                                                        & \multicolumn{1}{c|}{\multirow{-3}{*}{\cellcolor[HTML]{C0C0C0}\textbf{Spearman}}} & \textbf{avg} & \cellcolor[HTML]{FFFFFF}60           & \cellcolor[HTML]{FFFFFF}60             & \cellcolor[HTML]{FFFFFF}\textbf{100}  \\ \hline
		\multicolumn{1}{|c|}{\cellcolor[HTML]{C0C0C0}}                                                                                              & \multicolumn{2}{c|}{\cellcolor[HTML]{C0C0C0}\textbf{Pearson}}                    & 45           & 71                                   & \textbf{85}                                                                    \\ \cline{2-6}
		\multicolumn{1}{|c|}{\multirow{-2}{*}{\cellcolor[HTML]{C0C0C0}\textbf{\begin{tabular}[c]{@{}c@{}}Average \\ (leave-one-out)\end{tabular}}}} & \multicolumn{2}{c|}{\cellcolor[HTML]{C0C0C0}\textbf{Spearman}}                   & 44           & 54                                   & \textbf{74}                                                                    \\ \hline
	\end{tabular}
\end{table}

Having answered our research questions, we looked for correlations between ratings and typical measures used for evaluating AMT systems. In particular, we adopted the widely-used measures available in a Python package\cite{raffel2014mireval}. In Fig. \ref{fig:average}, the red line represents the F1-score computed considering as matches notes whose parameters lie within a certain range around the true value; in this case, we considered as parameters
\begin{inparaenum}
	\item the onset and offset times with a range of $\pm$ 50ms,
	\item the linearly re-scaled velocity so that the L2 error is minimized with tolerance of 10\%, and
	\item the pitch with tolerance of 1 quarter-tone\cite{hawthorne2018onsets}.
\end{inparaenum}
This measure, hereafter denoted as \textit{OBJ}, demonstrated low correlation with subjective ratings, mainly due to the following factors:

\begin{itemize}
	\item excerpt 0 has a low OBJ rating for O\&F (almost 0); however, subjects rated it much higher than O\&F,
	\item in the restoration task, O\&F received higher ratings than HR, which always has OBJ equal to 1 -- see Sec. \ref{sec:results}, and,
	\item occasionally, O\&F was rated lower than NR, which always has OBJ equal to 0.
\end{itemize}

Another interesting measure, named \textit{PEAMT}, reflecting subjective ratings was proposed in \cite{ycart2020investigating}. We computed Pearson and Spearman correlation coefficients between OBJ and PEAMT measures and the median and average values of the collected ratings. It was discovered that the average generally presents slightly higher correlation than the median, therefore the former was used for subsequent analyses. Table \ref{tab:correlations} shows that PEAMT correlates more strongly than OBJ to subjective ratings, especially for the Pearson coefficient. 
Nevertheless, especially in the \textit{restoration} task, correlation remains poor, motivating the search for an alternative measure.

We considered the features already used for the excerpt selection phase -- see Sec.~\ref{sec:chosing} -- except for audio-based features to be consistent with the existing evaluation methodologies.
We computed BPMs using the MIDI representation by counting how many onsets were present in sliding windows of size $0.1$, $1.0$, and $10$~s with a hop-size of 50\%. More precisely, we counted the mean and standard deviation for each window size. Importantly, to improve the portability of our measure, the features were first standardized using parameters computed on a large set of piano MIDI created by extracting piano solo performances from various datasets using \textit{ASMD}\cite{simonetta2020automatic}. After standardization, features of the predicted performances were subtracted from the target and the OBJ measure was appended. We performed linear regression on the dataset that we collected using various methods: Bayesian Ridge, \added{Automatic Relevance Determination}, Lasso Lars, Lasso, ElasticNet, Ridge and basic Linear regression. ElasticNet provided the best performance in terms of average L1 error.
To further improve the generalization ability, we trained a model using ElasticNet while removing features with low weights, i.e. $<$ 0.1.

We finally measured the average L1 error in a leave-one-out experiment, which provided $0.12$ for our measure, $0.19$ for PEAMT and $0.34$ for OBJ. Table \ref{tab:correlations} (bottom rows) shows correlation coefficients for each task and measure. 

When comparing PEAMT and our measure, one needs to consider differences in the design of the related tests. PEAMT is based on a test using audio clips lasting $5-10$~s, while, following the discussion summarized in \ref{sec:chosing}, this work employs clips lasting $20$~s. PEAMT authors' created 150 questions and collected 4 answers for each one; we instead preferred to collect more than 20 answers per question for plurality, while considering a control group which led us to reduce to 15 the total number of questions. At the same time, the space of possible note combinations is covered optimally (see Sec.~\ref{sec:chosing}). Furthermore, PEAMT is based on categorical questions -- subjects could chose between two systems -- with no HR and NR, while we measured a linear variable and included hidden and negative references. Finally, we focused on changing the context of the recordings and synthesis, but we included in our Transcription task the scenario used by PEAMT.

In general, we can state that PEAMT results agree with ours in finding low correlations between subjective ratings and OBJ, and that the two evaluation measures that we have built are rather similar in our preliminary tests. However, our test highlights new aspects that we think fundamental for audio restoration and that only our measure is able to tackle.

\section{Conclusion}
\label{sec:conclusions}
After conducting a thoroughly designed perceptual test, this work proposed a new approach for audio restoration: in the light of recent developments in audio signal processing, it becomes imaginable to recreate performances in the real world or through virtual instruments. We have therefore designed a perceptual test to assess to which extent existing technologies allow for such a methodology. It was discovered that the main limit lies in the usage of the MIDI format itself. Nonetheless, we proposed a new evaluation measure that seems consistent with the perception of context changes.

In case Standard MIDI Format is used as basis for the resynthesis, knowledge regarding contextual factors is required. Consequently, we argue that the future challenge for resynthesis-based audio restoration is in the conversion of the existing audio and music score in a new restored audio without the use of mid-level representations such as MIDI.


In this work, we have also identified limits for score-informed AMT, that, despite exploiting more information, lacks an effective feature fusion stage. Audio-to-score alignment should therefore become a main challenge for score-informed AMT; overcoming this problem could lead to improvements related to the exact knowledge of pitches and timings, leaving space for focusing on other parameters. Finally, we proposed a generic method to meaningfully choose excerpts when conducting music listening experiments.

\added{Future works include further experiments to assess the proposed experimental strategy and to confirm the presented results, as well as the development of new technologies for feature fusion and context-aware music transcription, such as Paraconsistent Feature Engineering~\cite{capobianco2019paraconsistent}, Wavelet transforms, and novel audio-to-score alignment methods~\cite{simonetta2021audiotoscore}}

\added{
	\section*{Reproducibility and Supplementary Information}
	To the sake of reproducibility, the whole code used for excerpt creations, answer collection, and statistical analysis is available online at the web address \url{https://github.com/LIMUNIMI/PerceptualEvaluation}.

	Two Supplementary files are also provided:
	\begin{enumerate}
		\item \texttt{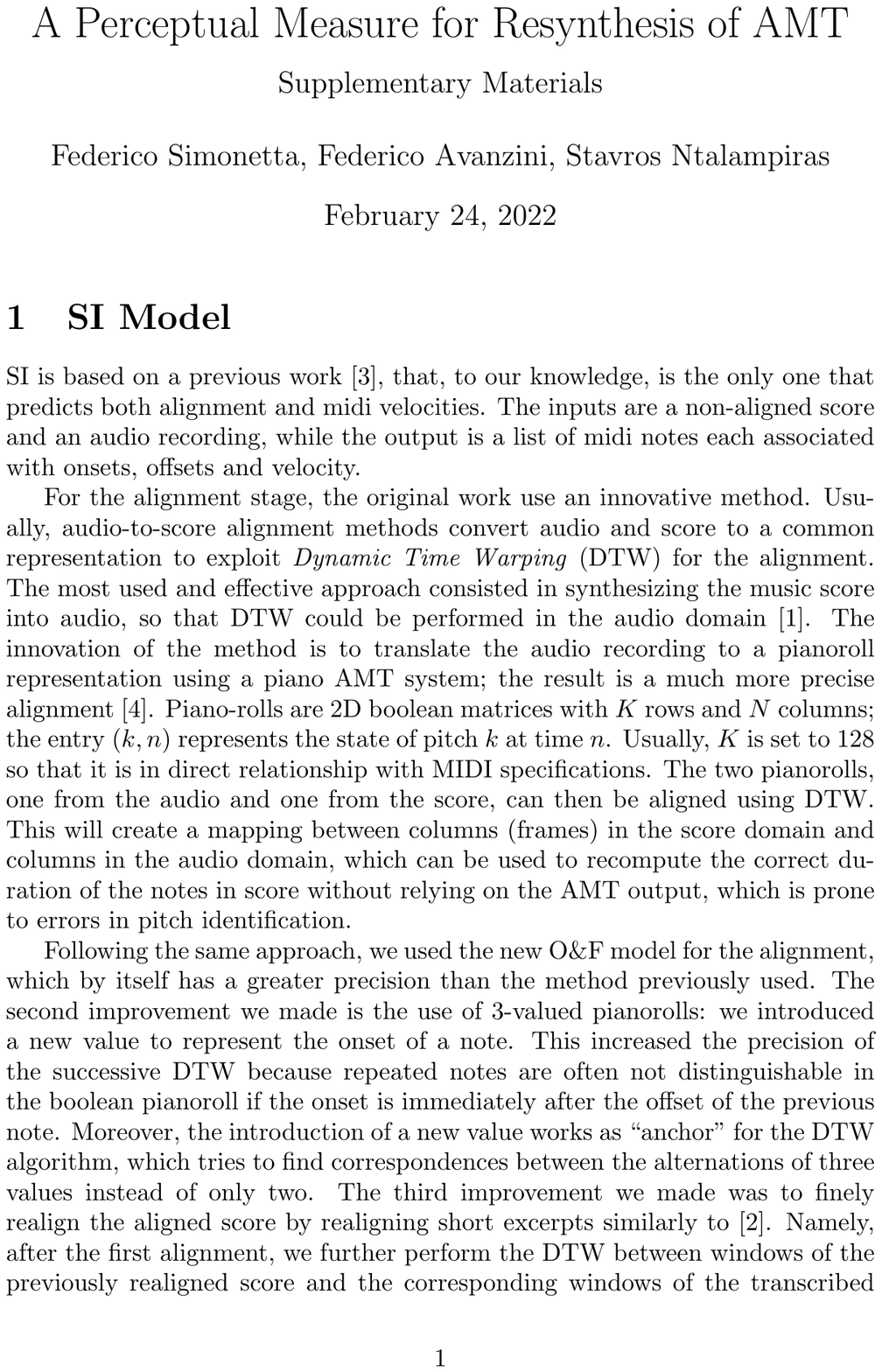}: contains the detailed description of the SI method used in this work -- see Sec.~\ref{sec:models};

		\item \texttt{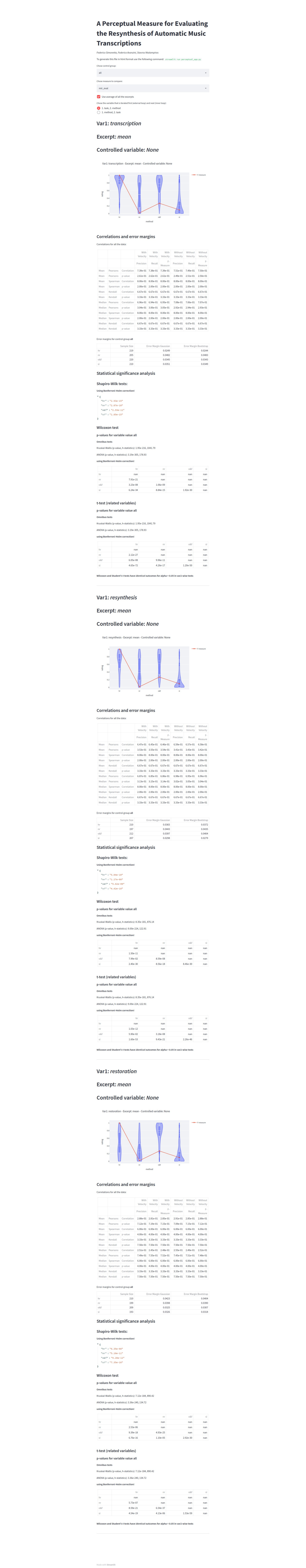}: contains extensive screenshots of the statistical analysis report used in this work. All the screenshots are generated using the code made available at the above URL. Specifically:

		      \begin{itemize}

			      \item the analysis of all responses -- no control groups -- per each task, averaged across the excerpts (page 1) and not (page 2);

			      \item the analysis of the expertise control groups, averaged across the excerpts (page 3) and not (page 4);

			      \item the analysis of the listening habits control groups, averaged across the excerpts (page 5) and not (page 6).

		      \end{itemize}
	\end{enumerate}
}

\section*{Acknowledgment}
We greatly acknowledge the support of Pianoteq for providing  a license of the physical modeled virtual piano; NVIDIA Corporation for the donation of a Titan V GPU; Julia Project, Python Software Foundation, and Linux Foundation for their invaluable work.

\newpage

\balance
\bibliographystyle{spmpsci}
\bibliography{./bibliography.bib}

\begin{thebibliography}{10}
\providecommand{\url}[1]{{#1}}
\providecommand{\urlprefix}{URL }
\expandafter\ifx\csname urlstyle\endcsname\relax
  \providecommand{\doi}[1]{DOI~\discretionary{}{}{}#1}\else
  \providecommand{\doi}{DOI~\discretionary{}{}{}\begingroup
  \urlstyle{rm}\Url}\fi

\bibitem{replaygain}
Replaygain 1.0 specification.
\newblock
  \urlprefix\url{http://wiki.hydrogenaud.io/index.php?title=ReplayGain_1.0_specification}

\bibitem{akbari2015piano}
{Akbari}, M., {Cheng}, H.: Real-time piano music transcription based on
  computer vision.
\newblock IEEE Trans. on Multimedia \textbf{17}(12), 2113--2121 (2015)

\bibitem{alias2016review}
Alías, F., Socoró, J., Sevillano, X.: A review of physical and perceptual
  feature extraction techniques for speech, music and environmental sounds.
\newblock Applied Sciences  (2016)

\bibitem{benetos201901automatic}
Benetos, E., Dixon, S., Duan, Z., Ewert, S.: Automatic music transcription: An
  overview.
\newblock {IEEE} Sig. Proc. Magazine \textbf{36}(1) (2019)

\bibitem{bigand2005multidimensional}
Bigand, E., Vieillard, S., Madurell, F., Marozeau, J., Dacquet, A.:
  Multidimensional scaling of emotional responses to music: The effect of
  musical expertise and of the duration of the excerpts.
\newblock Cognition and Emotion  (2005)

\bibitem{bogdanov2013essentia}
Bogdanov, D., Serra, X., Wack, N., Gómez, E., Gulati, S., Herrera, P., Mayor,
  O., Roma, G., Salamon, J., Zapata, J.: Essentia: an open-source library for
  sound and music analysis.
\newblock ACM Int. Conf. on Multimedia  (2013)

\bibitem{bolzinger1994study}
Bolzinger, S., Warusfel, O., Kahle, E.: A study of the influence of room
  acoustics on piano performance.
\newblock Journal De Physique Iv \textbf{4} (1994)

\bibitem{breebaart2017correlation}
Breebaart, J.: No correlation between headphone frequency response and retail
  price.
\newblock The Journal of the Acoustical Society of America \textbf{141}(66),
  EL526--EL530 (2017)

\bibitem{chernick2011bootstrap}
Chernick, M.R., González-Manteiga, W., Crujeiras, R.M., Barrios, E.B.:
  Bootstrap Methods, pp. 169--174.
\newblock Springer Berlin Heidelberg (2011)

\bibitem{contardo2020decremental}
Contardo, C.: Decremental clustering for the solution of p-dispersion problems
  to proven optimality.
\newblock INFORMS Journal on Optimization  (2020)

\bibitem{dannenberg2006interpretation}
Dannenberg, R.B.: The interpretation of {MIDI} velocity.
\newblock ICMC  (2006)

\bibitem{davies2001interpretation}
Davies, S., Sadie, S.: Interpretation.
\newblock Grove Music Online  (2001).
\newblock
  \urlprefix\url{https://doi.org/10.1093/gmo/9781561592630.article.13863}

\bibitem{devaney2017evaluation}
Devaney, J., Mandel, M.I.: An evaluation of score-informed methods for
  estimating fundamental frequency and power from polyphonic audio.
\newblock ICASSP  (2017)

\bibitem{dreyfus2020beyond}
Dreyfus, L.: Beyond the interpretation of music.
\newblock Journal of Musicological Research  (2020).
\newblock \doi{10.1080/01411896.2020.1775087}.
\newblock \urlprefix\url{https://doi.org/10.1080/01411896.2020.1714442}

\bibitem{erkut1990discrete}
Erkut, E.: The discrete p-dispersion problem.
\newblock European Journal of Operational Research  (1990)

\bibitem{everitt2011hierarchical}
Everitt, B.S., Landau, S., Leese, M., Stahl, D.: Hierarchical Clustering,
  chap.~4 (2011)

\bibitem{feiten2005audio}
{Feiten}, B., {Wolf}, I., {Eunmi Oh}, {Jeongil Seo}, {Hae-Kwang Kim}: Audio
  adaptation according to usage environment and perceptual quality metrics.
\newblock IEEE Trans. on Multimedia \textbf{7}(3), 446--453 (2005)

\bibitem{fu2011Trans.}
{Fu}, Z., {Lu}, G., {Ting}, K.M., {Zhang}, D.: A survey of audio-based music
  classification and annotation.
\newblock IEEE Trans. on Multimedia \textbf{13}(2), 303--319 (2011)

\bibitem{gabrielsson1999performance}
Gabrielsson, A.: The Performance of Music, second edition edn., chap.~14, p.
  501–602.
\newblock Cognition and Perception. Academic Press (1999)

\bibitem{gari2019analysis}
Garí, S.V.A., Kob, M., Lokki, T.: Analysis of trumpet performance adjustments
  due to room acoustics (2019)

\bibitem{geringer2007effects}
Geringer, J.M., Johnson, C.M.: Effects of excerpt duration, tempo, and
  performance level on musicians ratings of wind band performances.
\newblock Journal of Research in Music Education  (2007)

\bibitem{godsill1998digital}
Godsill, S.J., Rayner, P.J.W.: Digital Audio Restoration.
\newblock Springer London (1998)

\bibitem{goebl1999vienna}
Goebl, W.: The vienna 4x22 piano corpus (1999).
\newblock \urlprefix\url{https://doi.org/10.21939/4X22}

\bibitem{capobianco2019paraconsistent}
Guido, R.C.: Paraconsistent feature engineering [lecture notes].
\newblock IEEE Signal Processing Magazine \textbf{36}(1), 154--158 (2019)

\bibitem{gutierrezparera2018perception}
Gutierrez-Parera, P., Lopez, J.J.: Perception of nonlinear distortion on
  emulation of frequency responses of headphones.
\newblock The Journal of the Acoustical Society of America \textbf{143}(44),
  2085--2088 (2018)

\bibitem{hawthorne2018onsets}
Hawthorne, C., Elsen, E., Song, J., Roberts, A., Simon, I., Raffel, C., Engel,
  J., Oore, S., Eck, D.: Onsets and frames: Dual-objective piano transcription.
\newblock ISMIR  (2018)

\bibitem{plotly}
Inc., P.T.: Collaborative data science (2015).
\newblock \urlprefix\url{https://plot.ly}

\bibitem{jeong2020noteintensity}
Jeong, D., Kwon, T., Nam, J.: Note-intensity estimation of piano recordings
  using coarsely aligned midi score.
\newblock JAES \textbf{68} (2020)

\bibitem{jeong2017intensity}
Jeong, D., Nam, J.: Note intensity estimation of piano recordings by
  score-informed nmf.
\newblock Int. Conf. on Semantic Audio  (2017)

\bibitem{jillings2015audio}
Jillings, N., Moffat, D., De~Man, B., Reiss, J.D.: Web {A}udio {E}valuation
  {T}ool: {A} browser-based listening test environment.
\newblock SMC  (2015)

\bibitem{kalkandjiev2015influenceb}
Kalkandjiev, Z.S.: The influence of room acoustics on solo music performances:
  an empirical investigation.
\newblock Ph.D. thesis, TU Berlin (2015)

\bibitem{kalkandjiev2015influence}
Kalkandjiev, Z.S., Weinzierl, S.: The influence of room acoustics on solo music
  performance: An experimental study.
\newblock Psychomusicology \textbf{25}(33), 195–207 (2015)

\bibitem{kato2015effect}
Kato, K., Ueno, K., Kawai, K.: Effect of room acoustics on musicians’
  performance. part ii: Audio analysis of the variations in performed sound
  signals.
\newblock Acta Acustica united with Acustica \textbf{101}(44), 743--759 (2015)

\bibitem{klapuri2004automatic}
Klapuri, A.P.: Automatic music transcription as we know it today.
\newblock Journal of New Music Research \textbf{33}(3) (2004)

\bibitem{kob2020effect}
Kob, M., Amengual~Garí, S.V., Schärer~Kalkandjiev, Z.: Room Effect on
  Musicians’ Performance, pp. 223--249.
\newblock Springer International Publishing (2020)

\bibitem{kwon2017audiotoscore}
Kwon, T., Jeong, D., Nam, J.: Audio-to-score alignment of piano music using
  rnn-based automatic music transcription.
\newblock SMC  (2017)

\bibitem{luizard2018singers}
Luizard, P., Brauer, E., Weinzierl, S., Bernardoni, N.H.: How singers adapt to
  room acoustical conditions (2018)

\bibitem{marinelli2020musical}
Marinelli, L., Lykartsis, A., Weinzierl, S., Saitis, C.: Musical dynamics
  classification with cnn and modulation spectra.
\newblock SMC  (2020)

\bibitem{mizumachi2017discussion}
Mizumachi, M., Yamamoto, R., Niyada, K.: Discussion on subjective
  characteristics of high resolution audio.
\newblock Journal of The Audio Engineering Society  (2017)

\bibitem{mortberg2007dithered}
M\"ortberg, J.E.: Is dithered truncation preferred over pure truncation at a
  bit depth of 16-bits when a digital re-quantization has been performed on a
  24-bit sound file? (2007)

\bibitem{napoles2009effect}
Napoles, J.: The effect of excerpt duration and music education emphasis on
  ratings of high quality children's choral performances.
\newblock Bulletin of the Council for Research in Music Education  (2009)

\bibitem{naylor1992laboratory}
Naylor, G.M.: A laboratory study of interactions between reverberation, tempo
  and musical synchronization.
\newblock Acta Acustica  (1992)

\bibitem{orcalli2001methodologies}
Orcalli, A.: On the methodologies of audio restoration.
\newblock Journal of New Music Research \textbf{30}(4) (2001).
\newblock
  \urlprefix\url{https://www.tandfonline.com/doi/abs/10.1076/jnmr.30.4.307.7496}

\bibitem{potovcan2020estetska}
Potocan, Z.: Aesthetic perception of the singing voice in relation to the
  acoustic conditions.
\newblock Ph.D. thesis, University of Ljubljana (2020)

\bibitem{raffel2014mireval}
Raffel, C., McFee, B., Humphrey, E.J., Salamon, J., Nieto, O., Liang, D.,
  Ellis, D.P.W.: Mir\_eval: A transparent implementation of common mir metrics.
\newblock ISMIR  (2014)

\bibitem{ravi1994heuristic}
Ravi, S.S., Rosenkrantz, D.J., Tayi, G.K.: Heuristic and special case
  algorithms for dispersion problems.
\newblock Operations Research  (1994)

\bibitem{repp2006recording}
Repp, R.: Recording quality ratings by music professionals.
\newblock In: ICMC. Michigan Publishing (2006)

\bibitem{rizzi2019instrument}
{Rizzi}, A., {Antonelli}, M., {Luzi}, M.: Instrument learning and sparse nmd
  for automatic polyphonic music transcription.
\newblock IEEE Trans. on Multimedia \textbf{19}(7), 1405--1415 (2017)

\bibitem{salvador2007toward}
Salvador, S., Chan, P.: Toward accurate dynamic time warping in linear time and
  space.
\newblock Intelligent Data Analysis  (2007)

\bibitem{schwarz2016effects}
Schwarz, D., Lemaitre, G., Aramaki, M., Kronland-Martinet, R.: Effects of test
  duration in subjective listening tests.
\newblock ICMC  (2016)

\bibitem{simonetta201901multimodal}
Simonetta, F., Ntalampiras, S., Avanzini, F.: {Multimodal Music Information
  Processing and Retrieval: Survey and Future Challenges}.
\newblock {Int. Work. on Multilayer Music Representation and Processing}
  (2019)

\bibitem{simonetta2020automatic}
Simonetta, F., Ntalampiras, S., Avanzini, F.: Asmd: an automatic framework for
  compiling multimodal datasets.
\newblock SMC  (2020)

\bibitem{simonetta2021audiotoscore}
Simonetta, F., Ntalampiras, S., Avanzini, F.: Audio-to-score alignment using
  deep automatic music transcription.
\newblock In: Proceeddings of the IEEE MMSP 2021 (2021)

\bibitem{storm1980establishment}
Storm, W.: The establishment of international re-recording standards.
\newblock Phonographic Bulletin  (1980)

\bibitem{tanur2011margin}
Tanur, J.M.: Margin of Error, pp. 765--765.
\newblock Springer Berlin Heidelberg (2011)

\bibitem{ternstrom1989longtime}
Ternström, S.: Long-time average spectrum characteristics of different choirs
  in different rooms.
\newblock Voice (UK) \textbf{2}, 55--77 (1989)

\bibitem{ueno2005experimental}
Ueno, K., Kanamori, T., Tachibana, H.: Experimental study on stage acoustics
  for ensemble performance in chamber music.
\newblock Acoustical Science and Technology \textbf{26}(44), 345–352 (2005)

\bibitem{ueno2010effect}
Ueno, K., Kato, K., Kawai, K.: Effect of room acoustics on musicians’
  performance. part i: Experimental investigation with a conceptual model.
\newblock Acta Acustica united with Acustica \textbf{96}(3333), 505--515 (2010)

\bibitem{ueno2005cognitive}
Ueno, K., Tachibana, H.: Cognitive modeling of musician’s perception in
  concert halls.
\newblock Acoustical Science and Technology \textbf{26}(22), 156–161 (2005)

\bibitem{vonbekesy1968feedback}
Von~Békésy, G.: Feedback phenomena between the stringed instrument and the
  musician.
\newblock Rockefeller University Press (1968)

\bibitem{wang201710identifying}
Wang, S., Ewert, S., Dixon, S.: Identifying missing and extra notes in piano
  recordings using score-informed dictionary learning.
\newblock {IEEE}/{ACM} Trans. on Audio, Speech, and Language Processing
  \textbf{25}(10) (2017-10)

\bibitem{wapnick2005effects}
Wapnick, J., Ryan, C., Campbell, L., Deek, P., Lemire, R., Darrow, A.A.:
  Effects of excerpt tempo and duration on musicians' ratings of high-level
  piano performances.
\newblock Journal of Research in Music Education  (2005)

\bibitem{weinzierl2018sound}
Weinzierl, S., Lepa, S., Schultz, F., Detzner, E., von Coler, H., Behler, G.:
  Sound power and timbre as cues for the dynamic strength of orchestral
  instruments.
\newblock The Journal of the Acoustical Society of America \textbf{144}(3)
  (2018)

\bibitem{williams2016effect}
Williams, M.: Effect of excerpt duration on adjudicator ratings of middle
  school band performances.
\newblock Research Perspectives in Music Education  (2016)

\bibitem{wu2019polyphonic}
{Wu}, Y., {Chen}, B., {Su}, L.: Polyphonic music transcription with semantic
  segmentation.
\newblock ICASSP  (2019)

\bibitem{xu201905transferring}
Xu, M., Wang, Z., Xia, G.G.: Transferring piano performance control across
  environments.
\newblock In: ICASSP 2019 - 2019 IEEE International Conference on Acoustics,
  Speech and Signal Processing (ICASSP), pp. 221--225 (2019)

\bibitem{ycart2020investigating}
Ycart, A., Liu, L., Benetos, E., Pearce, M.T.: Investigating the perceptual
  validity of evaluation metrics for automatic piano music transcription.
\newblock TISMIR  (2020)

\bibitem{zapata2014multifeature}
Zapata, J.R., Davies, M.E.P., Gómez, E.: Multi-feature beat tracking.
\newblock IEEE/ACM Trans. on Audio, Speech, and Language Processing  (2014)

\end{thebibliography}

\end{document}